\documentclass{article}
\usepackage{spconf,amsmath,graphicx}
\usepackage{amsfonts}
\usepackage{amssymb}
\usepackage{tikz}
\usepackage{verbatim}
\usepackage{hyperref}
\usepackage{microtype}
\usepackage{chronology}
\usepackage{subfigure}
\usepackage{algorithmic}
\def\x{{\mathbf x}}

\definecolor{pinegreen}{cmyk}{0.92,0,0.59,0.25}
\definecolor{royalblue}{cmyk}{1,0.50,0,0}
\definecolor{purple}{rgb}{1,0,1}
\definecolor{violet}{cmyk}{0.79,0.88,0,0}
\tikzstyle{cblue}=[circle, draw, thin,fill=cyan!20, scale=0.5]
\tikzstyle{cgreen}=[circle, draw, thin,fill=green!20, scale=0.45]
\tikzstyle{cred}=[circle, draw, thin,fill=red!20, scale=0.45]
\tikzstyle{qgre}=[rectangle, draw, thin,fill=green!20, scale=0.8]
\tikzstyle{rpath}=[ultra thick, red, opacity=0.4]
\tikzstyle{legend_isps}=[rectangle, rounded corners, thin, 
                       fill=gray!20, text=blue, draw]
                        
\tikzstyle{legend_overlay}=[rectangle, rounded corners, thin,
                           top color= white,bottom color=green!25,
                           minimum width=2.5cm, minimum height=0.8cm,
                           pinegreen]
\tikzstyle{legend_phytop}=[rectangle, rounded corners, thin,
                          top color= white,bottom color=cyan!25,
                          minimum width=2.5cm, minimum height=0.8cm,
                          royalblue]
\tikzstyle{legend_general}=[rectangle, rounded corners, thin,
                          top color= white,bottom color=lavander!25,
                          minimum width=2.5cm, minimum height=0.8cm,
                          violet]
\usetikzlibrary{shapes}
\title{Information extraction from large multi-layer social networks}
\name{Brandon Oselio, Alex Kulesza, Alfred Hero\thanks{This work was partially supported by ARO grant \#W911NF-12-1-0443. We are grateful to Qiaozhu Mei who provided the Twitter data stream through his API gardenhose level access.}}
\address{Department of Electrical Engineering and Computer Science, University of Michigan, Ann Arbor, USA}
\definecolor{pinegreen}{cmyk}{0.92,0,0.59,0.25}
\definecolor{royalblue}{cmyk}{1,0.50,0,0}
\definecolor{purple}{rgb}{1,0,1}
\definecolor{violet}{cmyk}{0.79,0.88,0,0}
\tikzstyle{cblue}=[circle, draw, thin,fill=cyan!20, scale=0.5]
\tikzstyle{cgreen}=[circle, draw, thin,fill=green!20, scale=0.45]
\tikzstyle{cred}=[circle, draw, thin,fill=red!20, scale=0.45]
\tikzstyle{qgre}=[rectangle, draw, thin,fill=green!20, scale=0.8]
\tikzstyle{rpath}=[ultra thick, red, opacity=0.4]
\tikzstyle{legend_isps}=[rectangle, rounded corners, thin, 
                       fill=gray!20, text=blue, draw]
                        
\tikzstyle{legend_overlay}=[rectangle, rounded corners, thin,
                           top color= white,bottom color=green!25,
                           minimum width=2.5cm, minimum height=0.8cm,
                           pinegreen]
\tikzstyle{legend_phytop}=[rectangle, rounded corners, thin,
                          top color= white,bottom color=cyan!25,
                          minimum width=2.5cm, minimum height=0.8cm,
                          royalblue]
\tikzstyle{legend_general}=[rectangle, rounded corners, thin,
                          top color= white,bottom color=lavander!25,
                          minimum width=2.5cm, minimum height=0.8cm,
                          violet]
                          
\begin{document}
%
\maketitle
\begin{abstract}
Social networks often encode community structure using multiple
distinct types of links between nodes.  In this paper we introduce a
novel method to extract information from such multi-layer networks,
where each type of link forms its own layer.  Using the concept of
Pareto optimality, community detection in this multi-layer setting is formulated as
a multiple criterion optimization problem.  We propose an
algorithm for finding an approximate Pareto frontier containing a
family of solutions.  The power of this approach is demonstrated on a
Twitter dataset, where the nodes are hashtags and the layers
correspond to (1) behavioral edges connecting pairs of hashtags whose
temporal profiles are similar and (2) relational edges connecting
pairs of hashtags that appear in the same tweets.
\end{abstract}
\section{Introduction}

Social networks have become rich sources of data for network analysis,
where objectives might include community detection, edge prediction,
node behavior prediction, and model inference. However, it has become
increasingly difficult to extract meaningful information from these
networks due to the explosion in both the volume of data collected and
the diversity of available data types.  In this paper we focus on
addressing the latter problem for the task of community detection;
specifically, we consider networks containing multiple layers of
interactions between nodes.

For many social network applications, measures of association between
pairs of nodes may be available along multiple dimensions. For
example, graph edges may be observed directly in the data, or they may
be inferred from actions of the agents in the network. We make the
distinction between \textit{relational} links that are observed
explicitly and \textit{behavioral} links that are inferred from
ancillary data describing node behavior. Examples of relational links
between users might include observed interactions over a period of
time, mutually established friendship connections, or email
sender-reciever relationships.  Likewise, behavioral links might be
drawn between users who post items with similar semantic content, like
the same bands or movies, or exhibit correlated activity over
time. Further, it is possible to have multiple types of relational and behavioral links; for instance, there
could be both a professional and personal social network over the same
set of users.  Networks with multiple distinct edge types 
have been called multi-layer \cite{MaRo11}, multi-level \cite{SnBa03},
multi-relational, or multiplex \cite{KiAsBa13} networks.


In a multi-layer network, each layer may have a unique topology.  The
simplest way to apply existing network analysis algorithms (which
generally assume homogeneous edges) is to ``flatten" the data, i.e.,
to combine all the different types of links into a single-layer
network. This can be accomplished in various ways, for instance, by
performing a logical AND or OR on the layer-specific adjacency
matrices, or by computing their weighted (and possibly thresholded)
average. However, this approach has many hidden pitfalls; for example,
if one of the layers is noisier than the others then it probably
should not receive equal consideration when attempting community detection.

A better strategy, we argue, is to directly analyze the multi-layer
networks without flattening.  To show how this can be done, we propose
a new method of community detection for multi-layer networks. Our
approach employs multi-objective optimization, taking into account
multiple layers of network structure, which is then used to find a community
partition. We show that this algorithm
can provide significantly better community detection than that
obtained by standard single-layer techniques. 

The paper proceeds as follows.  In Sec. \ref{sec:multi_layer} we
define multi-layer networks. In Sec. \ref{sec:alg} a 
Pareto optimality approach to multi-layer community detection is proposed, and in
Sec. \ref{sec:twitter} we apply the proposed approach to Twitter data.  Finally, we discuss related work in Sec. \ref{sec:related} and
give concluding remarks in Sec. \ref{sec:conclusion}.

\section{Multi-layer Networks}
\label{sec:multi_layer}
A multi-layer network $G = (\mathcal V, \mathcal E)$ consists of
vertices $\mathcal V = \{ v_1, \ldots, v_p \}$, common to all layers,
and edges $\mathcal E = (\mathcal E_1, \ldots, \mathcal E_M)$ in $M$
layers, where $\mathcal E_k$ is the edge set for layer $k$, and
$\mathcal E_k = \{e^k_{v_i v_j}; \quad v_i, v_j \in V \}$.  Each edge
is undirected, though extensions to the directed case are not
difficult.  The multi-layer degree of a node $i$ is $d^i \in
\mathbb{R}^{M}$, with each entry $[d^i]_k$ being the degree of node
$i$ on layer $k$.

The adjacency matrix and degree matrix are defined as usual for each
layer:
\begin{align}
[[A^k]]_{ij} &= e^k_{v_i v_j} &
D^k &= \text{diag}([d^1]_k, [d^2]_k, \ldots, [d^p]_k)
\end{align}
Note that $D^k$ is simply a $p \times p$ diagonal matrix with the
layer-specific node degrees on the diagonal.

\section{Community Detection via Multiobjective Optimization}
\label{sec:alg}

Many existing community detection algorithms involve optimization
\cite{Fo10}. Methods that fall into this category include spectral
algorithms, modularity methods, and methods that rely on statistical
inference, particularly those that try to maximize a likelihood
function.  It seems natural that a multi-layer generalization of such
algorithms might somehow combine the optimization objective functions
as applied to each individual layer; this is the basis of
multi-objective optimization.

More formally, let community structure in a network be described by a
node partition $C$, where $C(i) = k$ means that node $i$ is in part
$k$.  Single-objective optimization methods of community detection
seek to find the partition ${\mathrm {argmin}}_C f(C)$ that minimizes
an objective function $f$ (which depends internally on the network
structure). In the following we consider the two community case; more communities can be found by a recursive use of the algorithm.

Now consider a two-layer network, and let $f_1$ and $f_2$ be objective
functions for the two layers. One obvious way of combining the layers
would be to minimize the linear combination $\alpha f_1(C)+(1-\alpha)
f_2(C)$ over $C$, where $\alpha \in [0,1]$.  However, linear
combination may be restrictive, especially when the objective
functions are complex.  A more general approach is instead to 
seek the Pareto optimal solutions of the multi-objective
minimization problem:
\begin{equation}
\hat{C}=\mathrm{argmin}_C[f_1(C),f_2(C)]~.
\label{eq:multiobj}
\end{equation}
A solution to the multi-objective optimization problem
(\ref{eq:multiobj}) is said to be weakly Pareto optimal (or weakly
non-dominated) if it is not possible to decrease any objective
function without increasing some other objective function \cite{Eh08,
  Ya10}. More formally, a solution $C_1$ \textit{dominates} a solution
$C_2$ if $f_i(C_1) \le f_i(C_2)$ for every objective function $f_i$
and there exists some $j$ such that $f_j(C_1) < f_j(C_2)$. The first
Pareto front is the set of weakly non-dominated points.

Calculating an exact Pareto front is, in general, a challenging
task. The most popular approximate methods are genetic algorithms,
which employ biologically inspired heuristics to attempt to transform
randomly selected seed cases into solutions on the Pareto front using
propagation. More details can be found in \cite{DePrAg02,CaDe08} and
the references therein. One disadvantage to genetic approaches is that
they are not deterministic. Additionally, there is no guarantee that
any of the Pareto front will be correctly identified. Finally, most
genetic algorithms deal with real-valued decision variables, while the
community detection problem has a discrete decision space.

The alternative strategy employed in this paper is based on the
Kernighan-Lin node swapping technique \cite{KeLi70}. The objective is
to find solutions that are approximately Pareto optimal. If it is
possible to obtain a sample of solutions that are likely to be on or
near the front, these points can be sorted for non-domination very
quickly \cite{DePrAg02}. In this way, a large set of solutions is
filtered to find candidates that are potentially Pareto optimal and
worth further consideration. Figure \ref{Pareto_alg} shows the
proposed algorithm.


\renewcommand{\algorithmicrequire}{\textbf{Input:}}
\renewcommand{\algorithmicensure}{\textbf{Output:}}

\begin{figure}
\begin{algorithmic}
\REQUIRE $f_1$, $f_2$
\STATE Obtain optimum solutions $C_1^*, C_2^*$ for each layer
\STATE Initialize $C = C_1^*$
\REPEAT
\FOR{$i: C(i) \neq {C_2^*}(i)$}
\STATE $C^{new} \leftarrow C$, $C^{new}(i) \leftarrow {C_2^*}(i)$ 
\STATE $\mathrm{cost}(i) \leftarrow f_2(C^{new}) - f_2(C)$
\ENDFOR
\STATE $i^* \leftarrow  \mathrm{argmin}_i\ \mathrm{cost}(i)$
\STATE $C(i^*) \leftarrow C_2^*(i^*)$
\UNTIL $C = {C_2^*}$
\ENSURE non-dominated solution values taken by $C$
\end{algorithmic}
\caption{Proposed algorithm for Pareto front identification.}
\label{Pareto_alg}
\end{figure}

For community detection, the objective is to minimize the ratio-cut $f_k$ for each layer $k=1,2$:
\begin{align}
  f_k(C) &= \frac{1}{2} \sum_{k = 1}^2\frac{\mathrm{cut}(C)}{|\{i: C(i) = k\}|} \\
  \mathrm{cut}(C) &= \sum_{C(i) = 1, C(j) = 2}[A^k]_{ij}
  \label{eq:cut}
\end{align}
A relaxed version of this objective function can be solved by
performing an eigendecomposition on the Laplacian $L_i = D_i -
A_i$. More details can be found in \cite{Lu07}.

\section{Twitter Dataset}
\label{sec:twitter}
The proposed algorithm was applied to a month of data from Twitter. A
two-layer network on hashtags was developed using tweets from October
2012. The data was obtained from the Twitter stream API at gardenhose
level access, which corresponds to 10\% of all tweets over the
month. A list of hashtags and the users who tweeted them was created
for each day, as well as the volume (i.e., number of observed
occurrences) of each hashtag per day.

\begin{figure}[htb]

\begin{minipage}[b]{0.37\linewidth}
	\centerline{\includegraphics[width = 0.9\linewidth]{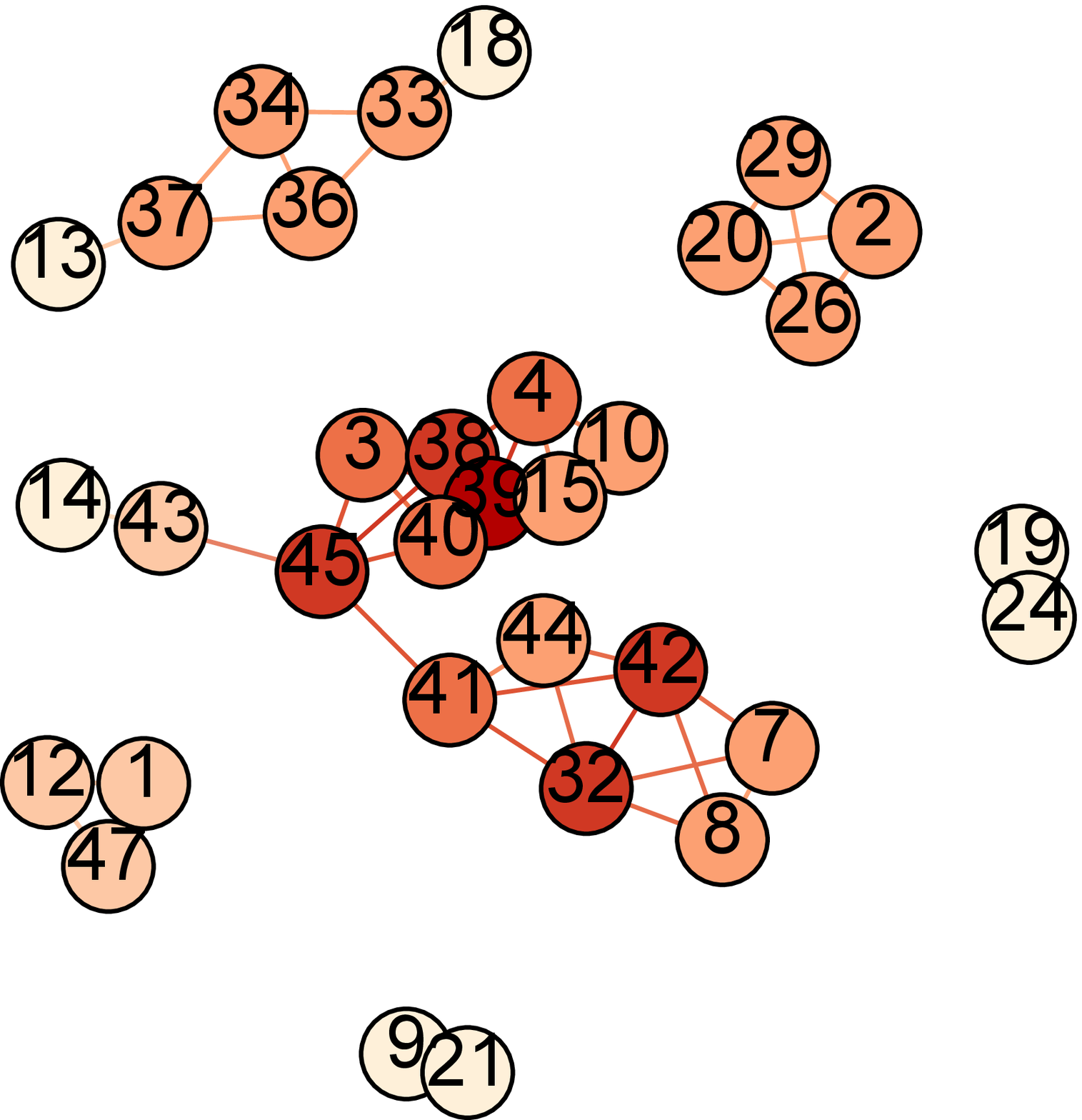}}
	\centerline{(a) Hashtag Volume Layer}
\end{minipage}
\hfill
\begin{minipage}[b]{0.5\linewidth}
	\centerline{\includegraphics[width = 0.9\linewidth]{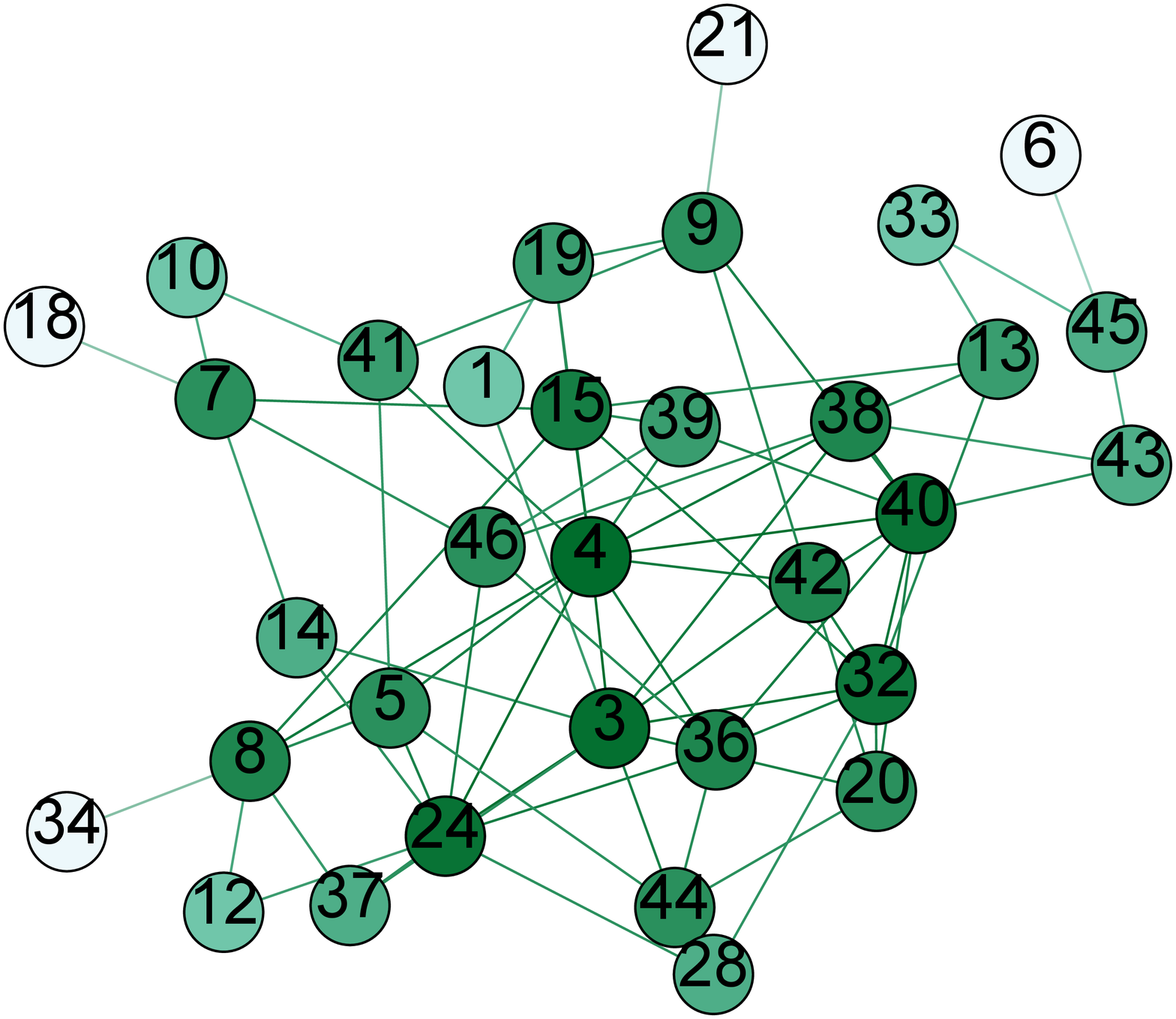}}
	\centerline{(b) Hashtag User Layer}
\end{minipage}
\caption{A network visualization of two layers of the hashtag dataset
  for October 10th, 2012. This example shows the differing topologies
  generated by different links in a network. While we see some
  similarities---for instance, nodes 38, 39, and 32 have high degree
  centralities in both networks---these networks have many
  differences, the most obvious being that the volume layer is not
  even fully connected, while the user layer is fully connected and
  has a diameter of only 6. }
\label{fig:network_view}
\end{figure}

Hashtags that were directly connected with the presidential election or politics were chosen out of a list of the most popular hashtags for the month, which yielded 48 hashtags. Figure \ref{fig:network_view} shows an example of two network layers for one day on the original set of 48 hashtags. In order to include some higher order connections, the list was expanded by including hashtags whose volume per day behaved similarly over the month as the first 48; this grew the network to 515 tags.


%

\begin{figure}[thb]
   \centerline{\includegraphics[width=0.95\linewidth]{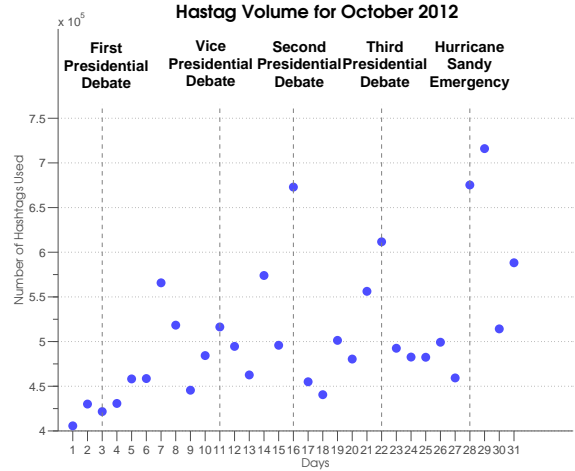}}
   \caption{Volume of observed usage of the 515 political hashtags
     along with an event timeline for October 2012. Notice that while
     we can see that some events correlate with
     hashtag usage for our dataset, this is not true for all events
     that might be expected to affect political hashtags.}
  \label{timeline}
\end{figure}

Initially, the total volume of the hashtags was studied over time, and real events were compared with the profile; this is shown in Figure \ref{timeline}. Some events are correlated with volume; Hurricane Sandy falls on the two day period with the largest hashtag volume. The second presidential debate also corresponds to a spike in hashtag volume. In contrast, the first presidential debate is not an identifiable event in the volume plot. 

\begin{figure}[htb]
\begin{minipage}[b]{0.47\linewidth}
\centering
\begin{tikzpicture}[
    every path/.style = {},
    every node/.append style = {font=\sffamily},
  scale = 0.36]
  
  \begin{scope}[shift = {(0, 6.5)}]
  \node (title1) at (10, 7) {\large User Layer};
  \tikzstyle{every node}=[ circle,minimum size=12pt,inner sep=0pt];
    \node[royalblue!20, fill,ellipse, text = black, draw] (N2) at (-0.7,1.5) {\footnotesize\#SystSci};
    \node[royalblue!20, fill,ellipse, text = black, draw] (N3) at (1.3,3.0) {\footnotesize\#DC};
    \node[royalblue!20, fill,ellipse, text = black, draw] (N4) at (0.8,5.6) {\footnotesize\#ICASSP2015};
    
    \draw[thin] (N2) -- (N4) node [ellipse, midway, above,sloped] (T1) {\tiny User 1};
    \draw[thin] (N4) -- (N3) node [ellipse, midway, above,sloped] (T1) {\tiny User 2};
    \end{scope}
     
    \begin{scope}[shift = {(0, -3)}]
  \node (title1) at (10, 8) {\large Volume Layer};
  \tikzstyle{every node}=[ circle,minimum size=12pt,inner sep=0pt];
    \node[royalblue!20, fill,ellipse, text = black, draw] (N2) at (-0.7,1.5) {\footnotesize\#SystSci};
    \node[royalblue!20, fill,ellipse, text = black, draw] (N3) at (1.3,3.0) {\footnotesize\#DC};
    \node[royalblue!20, fill,ellipse, text = black, draw] (N4) at (0.8,5.6) {\footnotesize\#ICASSP2015};
    
    \draw[thin] (N2) -- (N3) node [ellipse, midway, above,sloped] (T1) {};
    
    \end{scope}

          \node[text width = 4.5cm] (text1) at (14, 11) {User 1: \#ICASSP2015, \#SystSci};
     \node[text width = 4.5cm] (text1) at (14, 8.5) {User 2:  \#ICASSP2015, \#DC};   
   
 \draw[thin, dashed] (-4,6.5) -- (18, 6.5);
    \begin{scope}[shift = {(10,-1)}]

\draw[->] (0,0) -- (6,0) node[anchor=north] {time};
		(4,3.5) node{{\scriptsize Field weakening}};

\draw[->] (0,0) -- (0,4) node[anchor=east] {Volume};

\draw[thick, green] (0,0) -- (2,2) -- (4,2) -- (5,0);
\draw (1,3.5) node {\tiny\#ICASSP2015}; 

\draw[thick, green] (0,0.5) -- (2,1.5) -- (4,2) -- (5.3,0);
\draw (3,1.5) node {\tiny\#DC}; 

\draw[thick, red] (0,4) -- (1,0.5) -- (4,0.25) -- (5,1);
\draw[thick, green] (0,0.5) -- (2,1.5) -- (4,2) -- (5.3,0);
\draw (3,2.45) node {\tiny\#SystSci}; 
    \end{scope}
\end{tikzpicture}
\end{minipage}
  \caption{The two layers of the Twitter hashtag network are illustrated. At the top is the relational layer where a link between two hashtags indicates that at least one user used both hashtags in the same Tweet. At the bottom is the behavioral layer where a link indicates similarity in the hashtag usage volume over time. }
  \label{netwk_creation}
\end{figure}
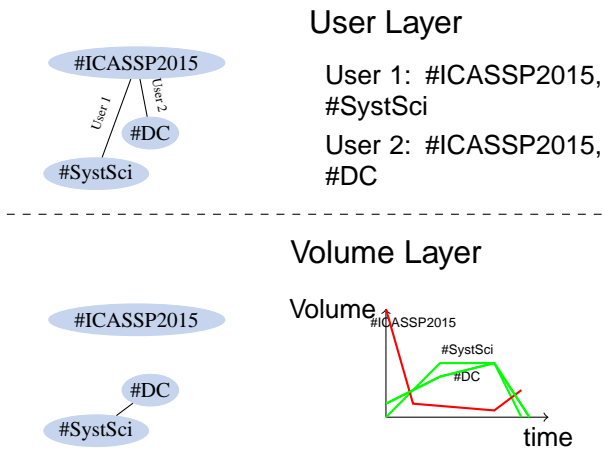

\begin{figure*}[htb]
\centering
\begin{minipage}[b]{.32\linewidth}
  \centering
  \centerline{\includegraphics[width=\linewidth]{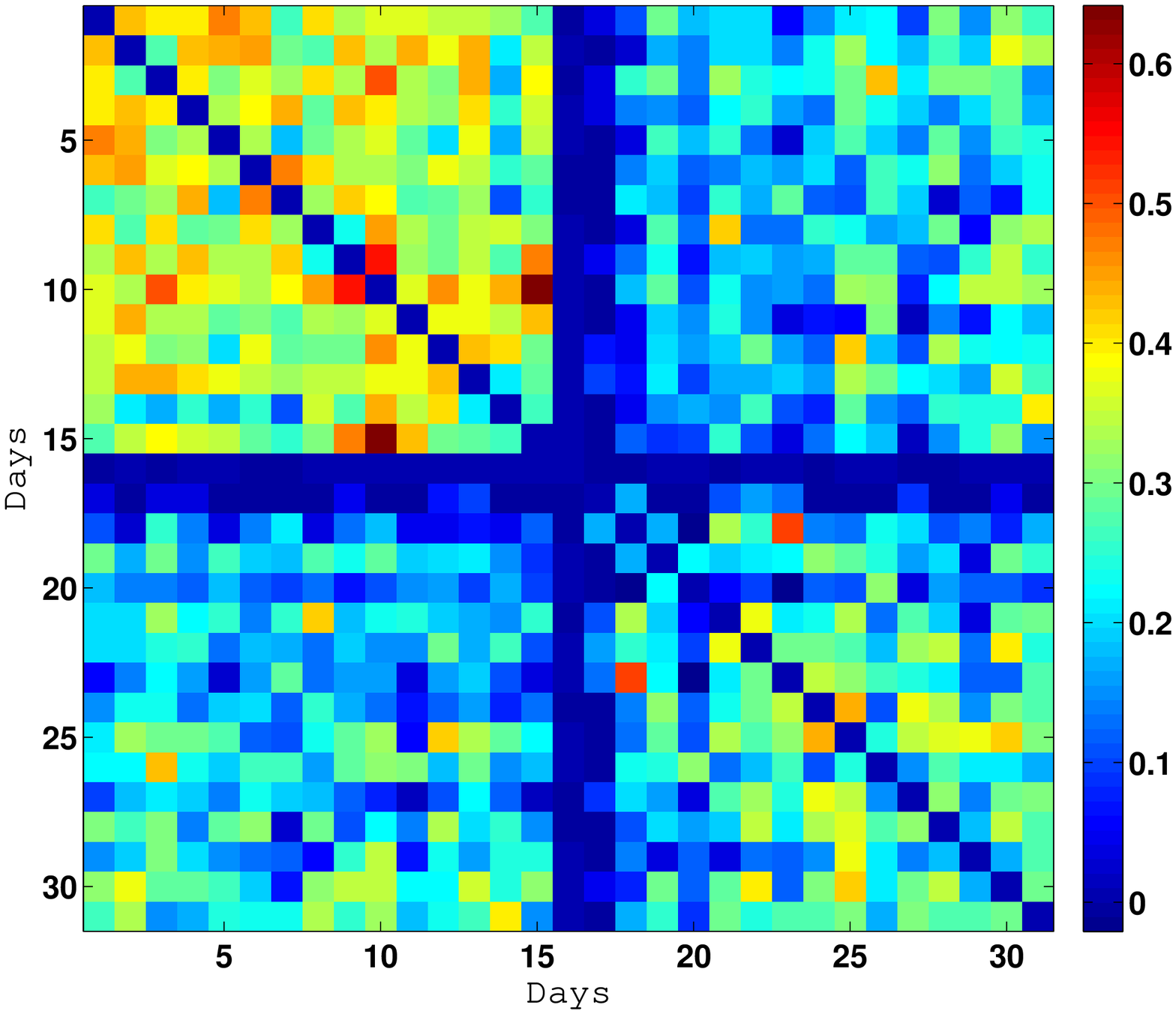}}
  \centerline{(a) Hashtag User Network}\medskip
\end{minipage}
\hfill
\begin{minipage}[b]{0.32\linewidth}
  \centering
  \centerline{\includegraphics[width=\linewidth]{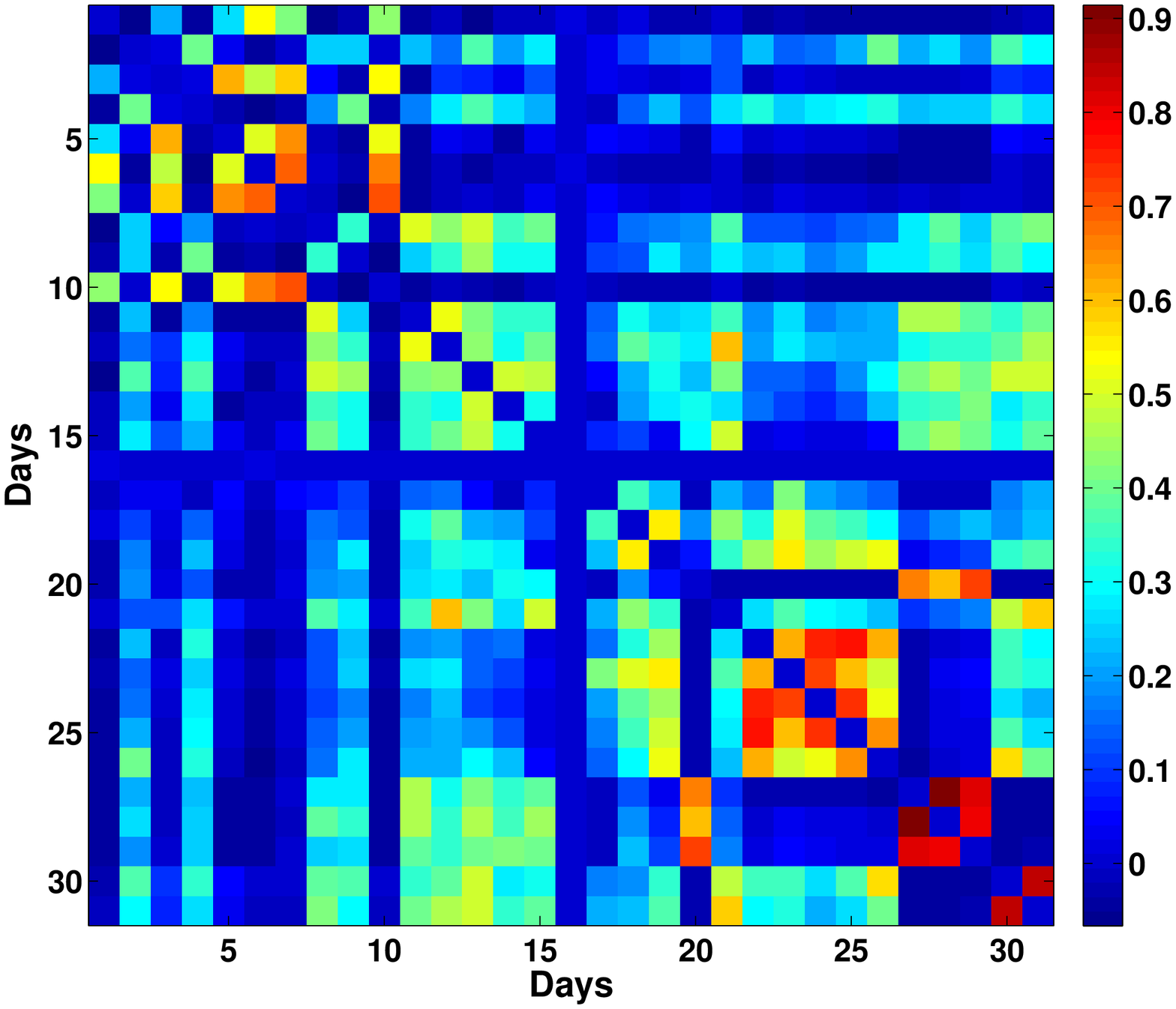}}
  \centerline{(b) Hashtag Volume Network}\medskip
\end{minipage}
\begin{minipage}[b]{0.32\linewidth}
  \centering
  \centerline{\includegraphics[width=\linewidth]{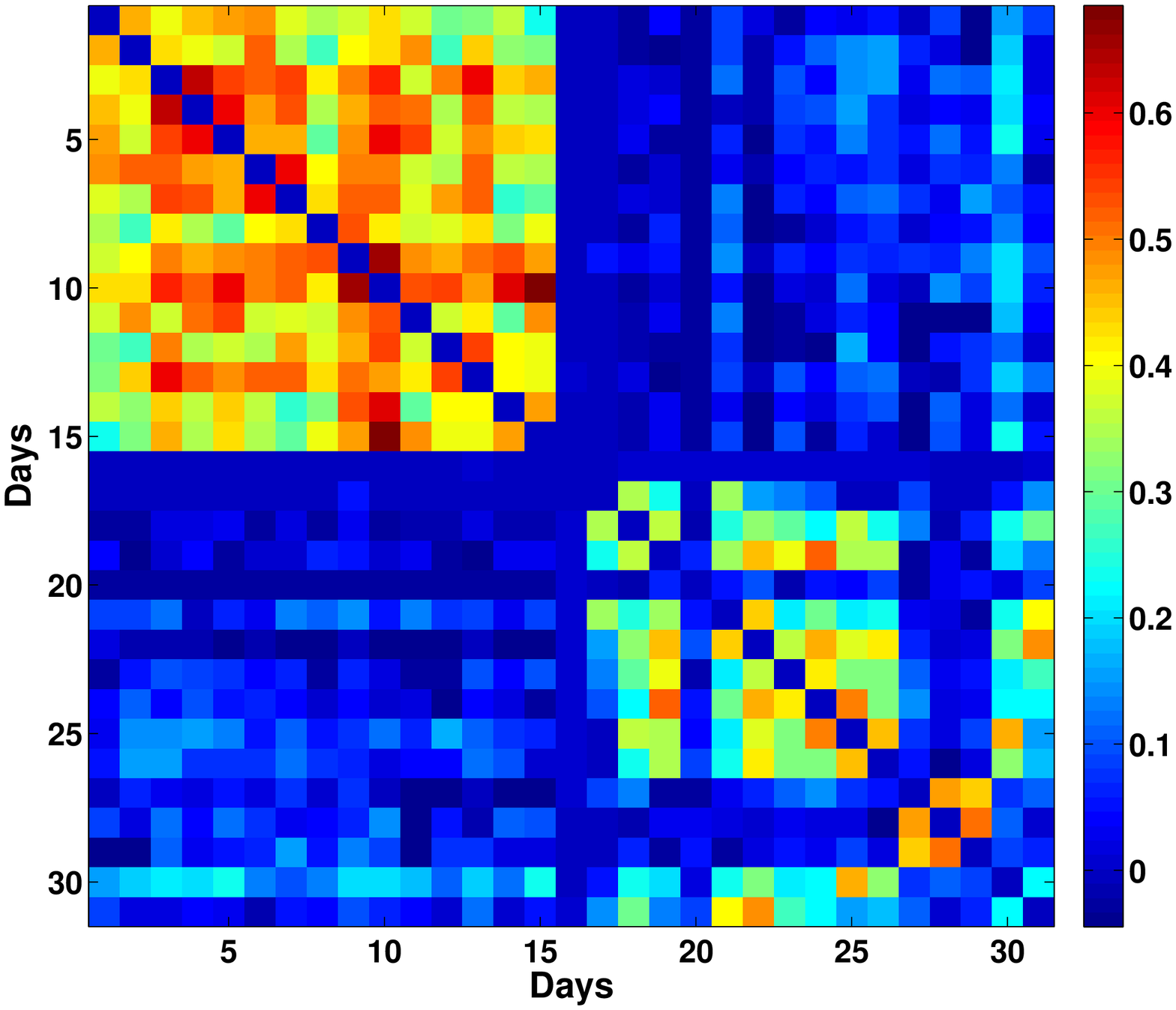}}
  \centerline{(c) Combined Network}\medskip
  \label{fig:user_com}
\end{minipage}

   \caption{The more highly resolved block structure in combined network heatmap clearly indicates that the hashtag community structure remains quite stable and coherent over the first 15 days of October but then breaks up into smaller clusters of coherency  over the remainder of the month. This may reflect the change of public opinions after the  second Presidential debates (October 16) and the effect of Hurricane Sandy (October 28) on Twitter hashtag volume and usage.   }
   \label{fig:user}
\end{figure*}

A time series of two-layer networks was created with hashtags as the nodes. Specifically, 31 two-layer networks were created by aggregating daily Tweet data over each day in the month. The first layer linked two hashtags if any user used both the hashtags in that particular day. This layer is referred to as the hashtag user layer. The second layer linked two hashtags if they had similar volume profiles over time. Intuitively, two hashtags would have a link with each other if they were popular or unpopular at the same time. So as not to take into account too much past data, the volume correlation was calculated using a moving window of 5 days. A Pearson correlation coefficient was used to calculate the correlations in volume for each pair of hashtags; the correlations then underwent a Fisher transformation and were thresholded by a value of 1.3859 which corresponds to an approximate 5\% false positive rate (in the bivariate normal case) when testing for the presence of a positive correlation \cite{Fi21}. This layer is referred to as the hashtag volume layer. Figure \ref{netwk_creation} demonstrates pictorially the creation of the two layers, using a simple dataset of three hashtags. 


We will show that one is able to obtain more information by the proposed Pareto multi-layer analysis methods than when the two layers are analyzed separately. To this end, the graph-cut partitions (\ref{eq:cut}) were computed for each day. We also computed approximately Pareto-optimal partitions by combining the single-layer solutions using Algorithm~\ref{Pareto_alg}, and selected a single partition by using the approximate midpoint of the Pareto front. The Adjusted Rand Index (ARI)~\cite{HuAr85} was then used to compare partitions on different days and see how hashtag relationships change over time. The ARI measures how similar partitions are, and can vary between -1 and 1.




Figure \ref{fig:user} shows heat maps of all the ARI indexes, both for the single layers considered separately as well as for the proposed algorithm. The hashtag user layer reflects fairly stable correlation among the two clusters until day 16, where there is a phase transition. Note that this phase transition also occurs on the volume layer heatmap. There is not much similarity between days in the user network, implying that there is not an optimal stable two cluster solution when considering the hashtag user layer alone, and it is difficult to extract real events.

In the hashtag volume layer heatmap, some community structure over days are highly correlated with each other. In particular, the days on which Hurricane Sandy occurs have communities that are highly correlated. It is also interesting to note that the communities at the end of the month are nothing like the bisected communities at the beginning, which implies considerable temporal evolution in the network. There is also more sparsity in the hashtag volume layer heatmap; consequently it may be possible to detect events more easily using this network.





The evident block structure in the Pareto combined heatmap shows that the multi-layer algorithm eliminates similarities between the first and second half of the months. The Pareto combined solution holds attributes from both the hashtag volume layer and hashtag user layer; the structural patterns that were present in the latter half of the month of the hashtag volume network are also present in the combined solution. The first half of the month also has some self-similarity, which is seen in the hashtag user layer. However, the proposed multi-layer algorithm was able to pick out some days that were more highly correlated than in either of the single layer solutions. In particular, days 3-5 are more highly correlated in the combined solution; October 3rd was the day of the first debate. Interestingly, the layers jointly reveal correlations between days not visible in the independent single layer analyses.

\section{Related Work}
\label{sec:related}
With the advent of large data, there has been more opportunity to explore this multi-layer structure. There has been some work in the modeling and representation of multi-layer networks, and how it relates to other studied problems \cite{DoSoCo13, KiAsBa13}. While there is a large body of work in single-layer community detection \cite{Fo10}, the multi-layer community detection literature is less comprehensive. Hypergraphs have been studied from a spectral perspective \cite{MiNa12}, which can be useful when dealing with a multi-layer structure. Some work in applying single-layer modularity methods to multi-layer structures is also available \cite{BaFaMa11}. For more information, see \cite{KiAsBa13}. This technique was also used in \cite{OsKuHe14}.

Multi-objective optimization has a long history \cite{CaDe08}. Here, we are only interested in a sorting algorithm used to find points that are possibly Pareto optimal; this is called non-dominated sorting. The method used in this paper is part of the evolutionary algorithm described in \cite{DePrAg02}. Some interesting application work has been done using multi-objective optimization \cite{JiSe08}, including supervised and unsupervised learning.

\section{Conclusion}
\label{sec:conclusion}
Multi-level network analysis is of growing interest as we are faced with increasingly complex data. In this paper, a method was introduced for finding communities in a multi-layer structure; it was demonstrated on a Twitter hashtag dataset and shown to deliver results that significantly differ from single layer analysis alone. The framework described can also be applied to other single-layer algorithms for the multi-layer setting. 

\bibliographystyle{IEEEbib}
\bibliography{ICASSP15}

\end{document}